\documentclass[12pt,a4paper]{article} 

\usepackage{graphicx}
\usepackage{xcolor}

\usepackage{caption}
\usepackage{subcaption}

\begin{document}

\title{Observational Constraints on the Structure-Induced Dark Energy Model}

\author{ A. Kaz\i m \c Caml\i bel
\\ \small Electrical and Electronics Engineering Department, 
\\ \small Turkish-German University, 34820 Beykoz, Istanbul, Turkey
\\ \small camlibel@tau.edu.tr}

\maketitle

\abstract{A new phenomenological dark energy model, originally associated to the large-scale structure formation and considered as a solution to the fine-tuning and coincidence problems related to the cosmological constant, was analyzed within the framework of General Relativity in a Friedman-Robertson-Walker spacetime and its model parameters were estimated using cosmic chronometers and recent DESI data. It turns out that the proposed model can serve as an alternative evolving dark energy model with a novel equation of state function, apart from other popular propositions in the literature. Due to the form of this phenomenological energy density ansatz, which starts to rise with the nonlinear structure growth in the universe and falls with the domination of cosmic voids, we prefer to call it structure-induced dark energy. Observational constraints show that it is not only a suitable solution for the fundamental problems such as coincidence or fine-tuning problems, it gives flexibility, when considering the cosmic tensions and presents a new perspective on the evolving dark energy models.
\\

}

\section{Introduction}

As is well known, introduction of cosmological constant, $\Lambda$, in the context of Einstein's static universe model, dates back to the early days of general relativity \cite{einstein1917kosmologische}. Yet, its prominent role as a constituent in the standart model of the universe started with the discovery of accelerated expansion of space \cite{riess1998observational, perlmutter1999measurements}. It was the most simplistic idea as a dark energy candidate and it was successful in explaining a variety of cosmological observations, such as CMB \cite{aghanim2020planck}, BAO \cite{anderson2014clustering} and Supernova Ia \cite{scolnic2022pantheon+}. 

Nevertheless, researchers started to propose alternative dark energy candidates, starting from the very first day, based on the fundamental and apparent problems with the $\Lambda$CDM model. Looking back at developments in the field, one can summarize objections to cosmological constant in three main categories; (i) fundamental problems with $\Lambda$, such as fine-tuning \cite{sahni2002cosmological} and coincidence problems \cite{zlatev1999quintessence}, (ii) cosmic tensions \cite{perivolaropoulos2022challenges} between the parameter estimations for $\Lambda$CDM at different redshifts ($H_0$ tension, $\sigma_8$ tension, etc.) and most recently (iii) evidence for evolving dark energy models with new observations \cite{adame2025desi}.

Alternative dark energy models were motivated by scalar fields (e.g. quintessence models) \cite{caldwell1998cosmological, tsujikawa2013quintessence}, or other types of fluids (e.g. Chaplygin gas \cite{kamenshchik2001alternative, bhattacharjee2025comprehensive}, barotropic fluids \cite{mandal2024late}, etc.). These models may propose solutions to coincidence and fine-tuning problems through tracking \cite{bag2018new} and thawing/freezing mechanisms \cite{caldwell2005limits} or elevate Hubble tension with early (EDE) \cite{kamionkowski2023hubble}, phenomenological emergent (PEDE) \cite{li2019simple} or transient dark energy models \cite{perkovic2019transient}. On the other hand, there are also phenomenological parametrizations that generalize dark energy with various ans\"atze either for energy density \cite{di2021dark} or equation of state functions, $w$ \cite{chevallier2001accelerating, linder2003exploring}. Such parametrizations, which allow further degrees of freedom for the model, gave rise to the possible exclusion of $\Lambda$CDM, considering new observations \cite{adame2025desi}; of course only for the accepted parameter space. It is yet early to discuss a full exclusion of $\Lambda$, at least alternative parametrizations of energy density or $w$ should be considered and analyzed first.

In a recent work \cite{ccamlibel2025cosmic} we proposed an evolving dark energy ansatz, which was initially inspired by the structure formation and possible binding energy of the dark matter structures. The corresponding ansatz provided a dark energy density, which starts at a certain redshift and grow from there with growing structure and increasing binding potential within dark matter structures in the universe. The main goal of this proposal was to offer a {\it natural} alternative to $\Lambda$ that is free of coincidence and fine-tuning problems and also may release the tensions from observations. This structure-induced dark energy density (SIDE) would reach a maximum and starts to decrease with the domination of cosmic voids. Corresponding model parameters were estimated using supernovae type Ia data and quasars. 

In this work, we treat SIDE model as a an alternative evolving dark energy density ansatz and first discuss its evolving properties within the scope of general relativity in the next chapter. Then we introduce two distinct types of data, cosmic chronometers and distance measures from baryon acoustic oscillations and compare SIDE model with them along with $\Lambda$CDM, in order to estimate the model parameters and to see how good SIDE represents the data. Additionally we compare the results with the popular $w_0-w_a$ ansatz and discuss our results.

\section{Characteristics of the SIDE Model}

\begin{figure}
\centering
\begin{subfigure}{.3\textwidth}
\centering
\includegraphics[width=1.1\linewidth]{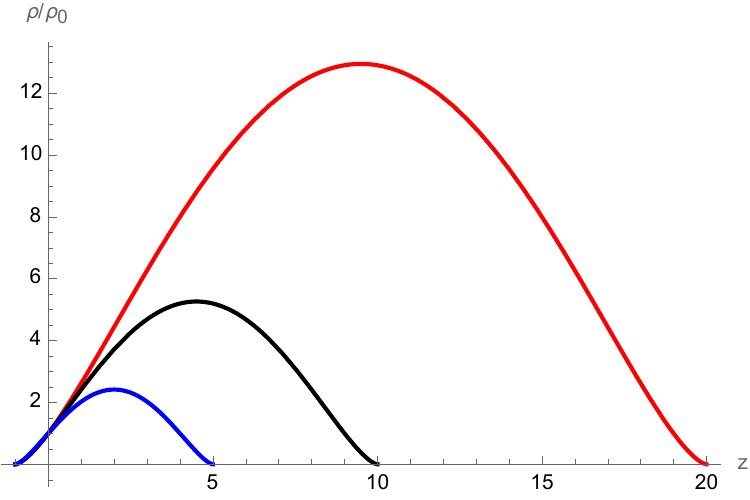}
\caption{}
\label{zstar_dep}
\end{subfigure}
\begin{subfigure}{.3\textwidth}
\centering
\includegraphics[width=1.1\linewidth]{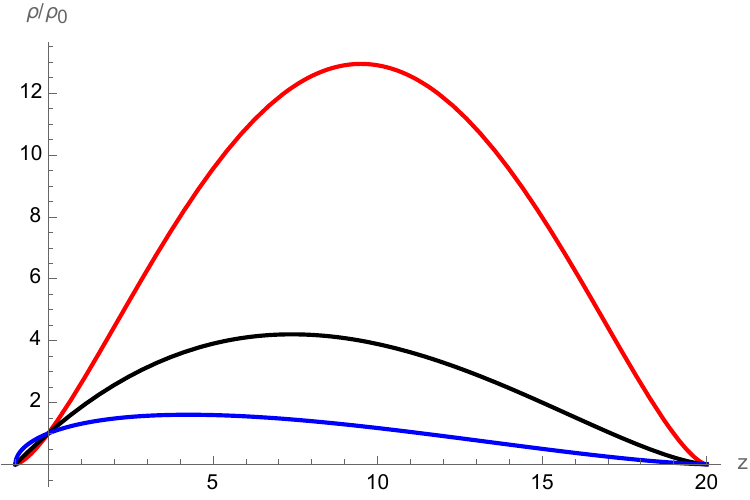}
\caption{}
\label{alpha_dep}
\end{subfigure}
\begin{subfigure}{.3\textwidth}
\centering
\includegraphics[width=1.1\linewidth]{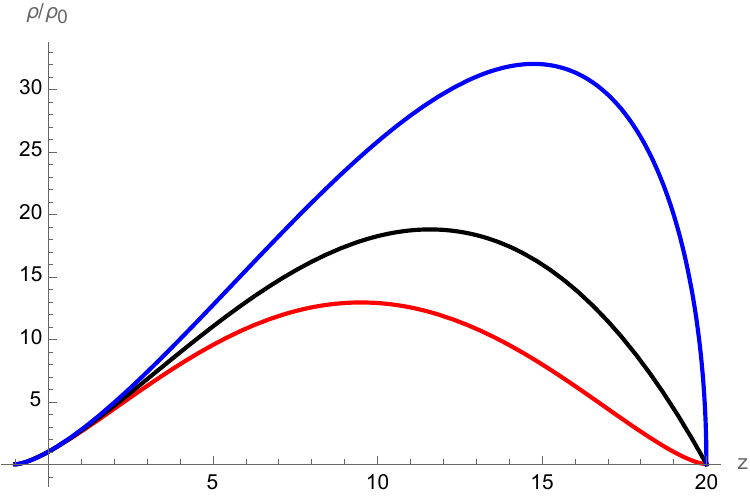}
\caption{}
\label{beta_dep}
\end{subfigure}
\caption{Plots showing the evolution of the SIDE density, $\rho$, for varying each three model parameters, consecutively, where the other two were kept constant, such that, (a) $z_* =$ 5 (blue), 10 (black), 20 (red) (b) $\alpha =$ 1.5 (red), 1 (black), 0.5 (blue) (c) $\beta =$ 1.5 (red), 1 (black), 0.5 (blue).}
\label{parameter-dep}
\end{figure}

The phenomenology for the SIDE can be summarized as a proposed energy density, which is initiated at a given redshift as a result of growth of inhomogeneities, increases as structure growth continues up to a certain maximum density value and after that starts to decay. The ansatz suggested for this behavior was,

\begin{equation}
\rho_{\rm SIDE}(z)=\rho_0(1+z)^{\alpha}(1-z/z_*)^{\beta}
\label{side_density}
\end{equation}
where $\alpha$ determines how fast the SIDE density would decay after reaching the maximum, $\beta$ gives the rate of increase in SIDE density and $z_*$ is the redshift at which SIDE emerges. In the upcoming chapters we focus on two independent values $z_*=10$ and $z_*=100$. In accordance with the explained phenomenology and the given ansatz, both $\alpha$ and $\beta$ are taken to be positive. One can see from Figure \ref{parameter-dep}, how the variations of the parameters $\alpha$, $\beta$ and $z_*$ affect the evolution of SIDE density. A change in $z_*$ shifts the starting point of SIDE and the maximum value it can reach, a decrease in $\alpha$ pushes the turn-over point towards later cosmological times and even to the future, a decrease in $\beta$ means a steeper increase in the past, such that $\beta<1$ is an abrupt emergence of SIDE density.

\begin{figure}
\centering
\begin{subfigure}{.45\textwidth}
\centering
\includegraphics[width=1\linewidth]{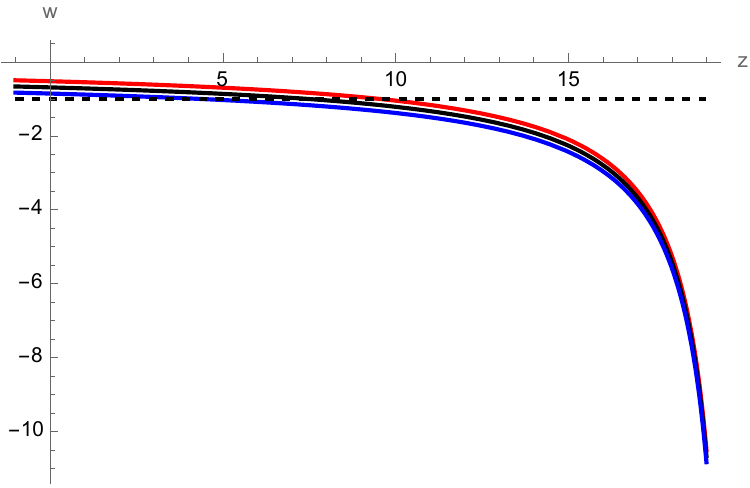}
\caption{}
\label{w_alpha}
\end{subfigure}
\begin{subfigure}{.45\textwidth}
\centering
\includegraphics[width=1\linewidth]{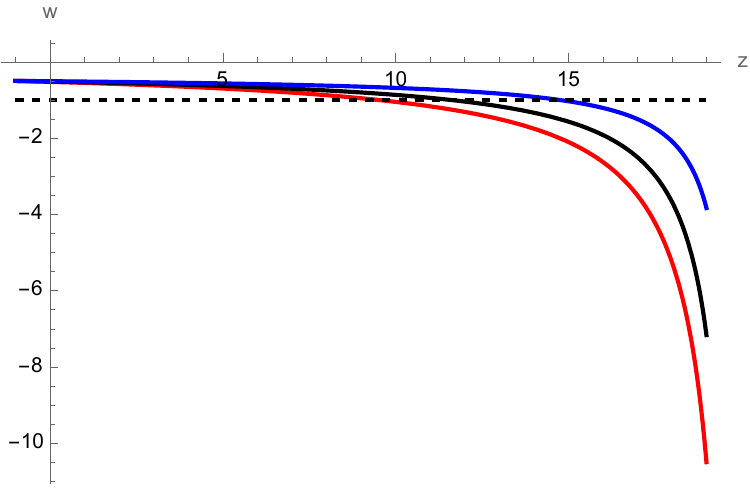}
\caption{}
\label{w_beta}
\end{subfigure}
\caption{Evolution of the equation of state parameter, $w$, for (a) $\alpha =$ 1.5 (red), 1 (black), 0.5 (blue) (b) $\beta =$ 1.5 (red), 1 (black), 0.5 (blue), together with $w=-1$ line (dashed). For each panel, the other parameter and $z_*$ were kept constant}
\label{w_dependence}
\end{figure}

To investigate the suggested model in terms of fluid pressure and equation of state function we assume a perfect fluid energy-momentum tensor under the framework of General Relativity within a Friedman-Robertson-Walker spacetime, which gives the conservation equation for any fluid component,

\begin{equation}
\frac{d}{dt}(a^3\rho)=-p\frac{d}{dt}(a^3)
\end{equation}
where $p$ is the pressure of the perfect fluid and $a$ is the scale factor, defined as $a\equiv1/(1+z)$. Rearranging the terms to get $p$ in terms of $\rho$ and $a$, we have

\begin{equation}
p(a)=-(3\rho+\rho'a)/3
\end{equation}
where prime denotes a derivative with respect to $a$. Putting the definition (\ref{side_density}) into the above equation gives,

\begin{equation}
p_{\rm SIDE}(z)=\rho_{\rm SIDE}(z)\left(-1+\frac{\alpha}{3}-\frac{\beta}{3}\frac{(1+z)}{(z_*-z)}\right)
\label{SIDE_p}
\end{equation}
where the expression in parentheses is the effective equation of state parameter, $w(z)$. It is easy to see that SIDE model acts as a three-parameter generalization of $\Lambda$ and a new type of parametrization for $w(z)$, distinct from other existing dynamical dark energy models. When $\beta$ is equal to 0, we have a dark fluid with constant $w$, whereas if both $\alpha$ and $\beta$ are negligible, we get a cosmological constant. Dependence of $w$ on varying model parameters can be seen in Figure \ref{w_dependence}.

\section{Data and Analysis}

\begin{figure}
\centering
\includegraphics{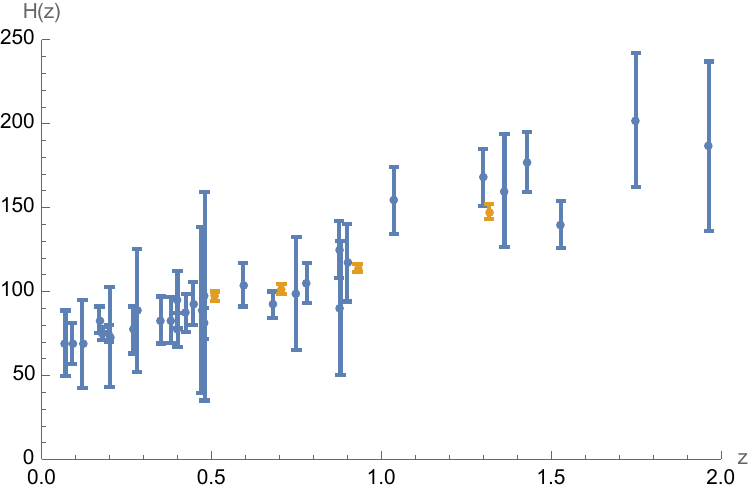}
\caption{$H(z)$ data derived from cosmic chronometers (blue) \cite{moresco2022unveiling} and from DESI DR1 distance data (gold) \cite{adame2025desi}.}
\label{CC_DESI_data}
\end{figure}

In order to estimate the model parameters and to see if SIDE model can explain the current cosmological observations, we combine to sets of data in a single Hubble diagram: Cosmic chronometers, and Hubble distance values from DESI DR1 (Figure \ref{CC_DESI_data}).

Cosmic chronometers are cosmological probes, which relate the age of the old cosmological objects (mainly galaxies) to the age of the universe and the Hubble rate $H(z)$ at the measured redshift of the observations. (For details and the data set used in this work see \cite{moresco2022unveiling})

The Hubble expansion rates from DESI DR1 \cite{adame2025desi}, are derived using the basic relation

\begin{equation}
D_{\rm{H}}(z)\equiv c/H(z)
\label{desi-equation}
\end{equation}
where the Hubble distance $D_{\rm{H}}(z)$ are measured from baryon acoustic oscillations in the form of $D_{\rm{H}}(z)/r_{\rm{d}}$ at five different redshifts, using luminous red galaxies, emission line galaxies and Lyman-$\alpha$ forest quasars. In order to obtain $H(z)$ from (\ref{desi-equation}), the value of the sound horizon at {\it drag} epoch, $r_{\rm{d}}=147.09\pm0.26$, is taken from CMB measurements \cite{aghanim2020planck}.

For the estimation of model parameters ($H_0$, $\Omega_{\rm M,0}$, $\alpha$, $\beta$) we choose two alternative subscenarios of SIDE, where we assume an initiation redshift for the dark energy density at $z_*=10$ (SIDE$_{10}$) and at $z_*=100$ (SIDE$_{100}$). We assume spatial flatness and use 

\begin{equation}
H(z)=H_0\sqrt{\Omega_{\rm{M},0}(1+z)^3+(1-\Omega_{\rm{M},0})(1+z)^{\alpha}(1-z/z_*)^{\beta}}
\label{fitting_ansatz}
\end{equation}
as our fitting ansatz.

Best-fit results, together with the fit for $\Lambda$CDM are summarized in Table \ref{parameters_CC_DESI}, along with AIC and BIC values for each fit.

AIC and BIC values hint that it is still more likely that analyzed data can be explained by $\Lambda$CDM model, rather than suggested SIDE models. Nevertheless there is still a statistically significant probability that the observations can be produced with SIDE. A slightly higher value of $\Omega_{\rm M,0}$ with respect to $\Lambda$CDM is expected, since SIDE model exhibits phantom behaviour, which should be compensated by a reduced need for a dark energy contribution. 

An increased number of parameters and their codependence result in higher uncertainties within SIDE. Estimated $H_0$ value covers variety of contradicting $H_0$ results from literature within 2-$\sigma$. Similarly, $\alpha$ and $\beta$ are consistent with  $\Lambda$CDM within 2-$\sigma$ confidence interval.

\begin{table}
\caption{Best fit results for cosmic chronometer and DESI DR1}
\centering
\begin{tabular}{@{}*{7}{l}}
\hline
& $H_0$ & $\Omega_{\rm M,0}$ &  $\alpha$ & $\beta$ & AIC & BIC \\
\hline
$\Lambda$CDM & 69.61$\pm$1.66 & 0.30$\pm$0.02 & - & - & 286.4 & 291.2 \\
\hline
SIDE$_{10}$ & 64.79$\pm$5.14 & 0.37$\pm$0.06 & 5.47$\pm$3.85 & 39.24$\pm$25.40 & 288.4 &296.5 \\
\hline
SIDE$_{100}$ & 64.41$\pm$5.20 & 0.38$\pm$0.06 & 6.50$\pm$4.45 & 480.94$\pm$307.42 & 288.3 & 296.4 \\
\hline
\end{tabular}
\label{parameters_CC_DESI}
\end{table}

\begin{figure}
\centering
\includegraphics{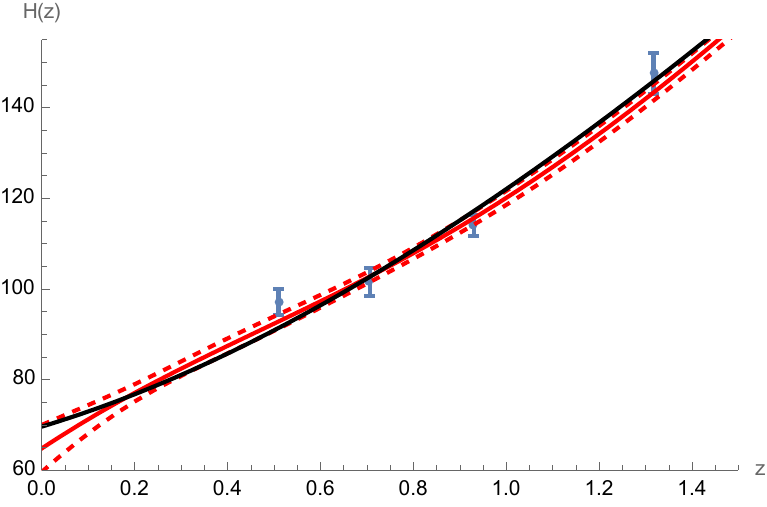}
\caption{Best fits for $\Lambda$CDM (blue) and SIDE$_{10}$ (red), with 1-$\sigma$ confidence intervals (dashed), together with DESI DR1 $H(z)$ datapoints.}
\label{DESI_lcdm_side102}
\end{figure}

In Figure \ref{DESI_lcdm_side102}, we show the best-fitting $\Lambda$CDM and SIDE$_{10}$ with DESI DR1 $H(z)$ datapoints. Both models and the datapoints agree with each other within 1$\sigma$ intervals.

\begin{figure}
\centering
\includegraphics{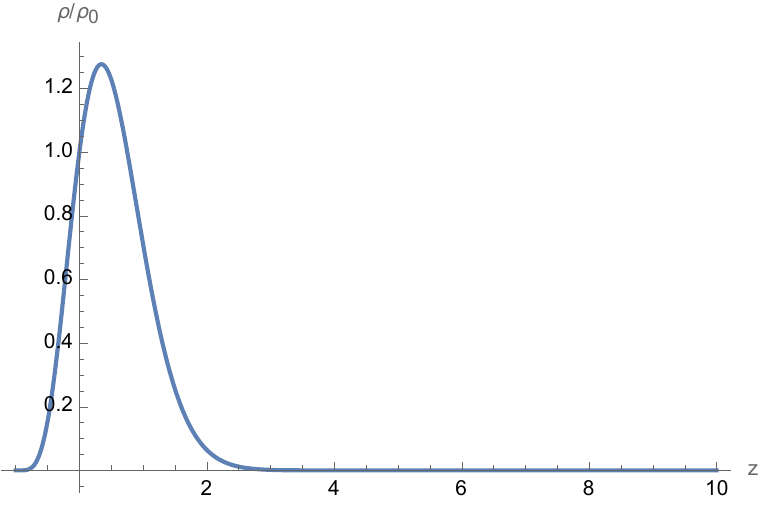}
\caption{Evolution of the energy density, $\rho$,  for the best fitting SIDE$_{10}$, when the cosmic chronometers and DESI DR1 considered together.}
\label{side102_rho}
\end{figure}

\begin{figure}
\centering
\includegraphics{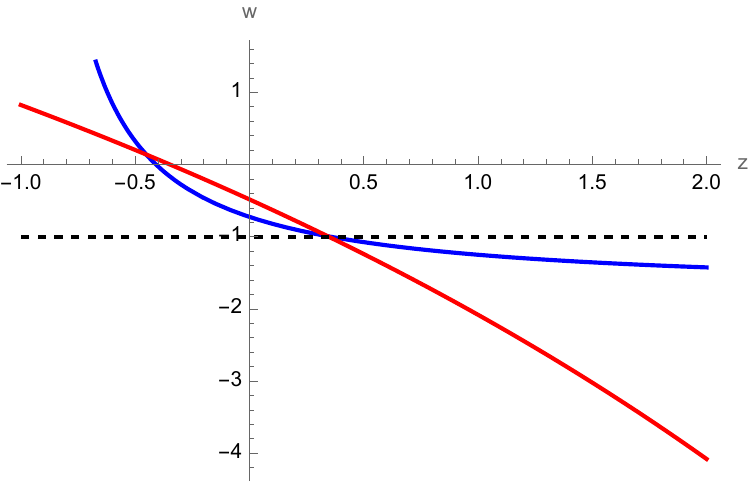}
\caption{Evolution of the equation of state parameter as a function of redshift, for SIDE$_{10}$ (red), $(w_0-w_a)$DE model (blue) \cite{adame2025desi} and for $\Lambda$ (black, dashed)}
\label{eos_all}
\end{figure}

With the estimated parameters, SIDE density reaches a maximum at around $z\simeq0.35$ (Figure \ref{side102_rho}). This behaviour indicates a crossing from phantom region towards a decaying one. Characteristics Of the SIDE density beyond that point can be investigated by examining the $w(z)$ using Equation \ref{SIDE_p}. In Figure \ref{eos_all}, we give $w(z)$ derived for SIDE$_{10}$ together with $\Lambda$ (horizontal line of $w=-1$) and best fit result for popular $(w_0-w_a)$ from \cite{adame2025desi}. Both dynamical models exhibit an increase in $w(z)$ towards the present day and allow positive $w$ values for the future. $(w_0-w_a)$ parametrization diverges beyond $z\sim-0.5$, whereas SIDE model shows a steady behaviour. It is interesting to see when both models plotted there are two triple points in the graph. First one is at the phantom-crossing ($w=-1$) around $z\sim0.35$ and the second one is at $z\sim-0.4$ where both curves cross $w=0$. We leave it for future consideration, if there is any fundamental significance of those points with regards to dynamical dark energy models in general. 

\section{Conclusion}

Motivation for the SIDE model was to come up with a phenomenological dark energy model that would be triggered by the formation of large scale structures and driven by the long-range dark matter self-interactions, which would start to die out after a certain point in the evolution of structures, possibly at a time, when cosmic voids dominate the universe. Beyond that, the proposed ansatz can be accepted as a general model for any dark energy density that starts at a given redshift, reaches a maximum and starts to decrease after that. Examining this ansatz in terms of general relativity, we were able to show that equation of state function for this model is a new alternative parametrization for the evolving dark energy models. 

Comparison of the SIDE model against the $H(z)$ data from cosmic chronometers and BAO, gave insights on the following points:

  \item[$\bullet$] Statistical analysis through AIC and BIC parameters shows that $\Lambda$CDM is still more probable than SIDE model, yet there is still nonnegligible possibility that observed $H(z)$ data can be a result from a SIDE-driven universe. That is mainly due to AIC/BIC penalizes the high number of model parameters, which seems inevitable for generalizations of $\Lambda$CDM, since it is still the most simplistic model as the standart model of cosmology.
  \item[$\bullet$] Considering the $\alpha$-$\beta$ parameter space, we see that cosmological constant is covered within 2-$\sigma$ confidence interval. This result hints that it is hard to exclude $\Lambda$ indefinitely, considering evolving dark energy models, since the results may vary depending on the form of generalization of dark energy. For example, recent observational constraints using $w_0-w_a$ parametrization results in an exclusion of $\Lambda$ by a 3-$\sigma$ confidence level \cite{abdul2025desi}, yet a different choice of parametrization can give different results.
  \item[$\bullet$] SIDE model was proposed as a solution to coincidence and fine-tuning problems of $\Lambda$ \cite{ccamlibel2025cosmic}. Additionally, through the $H(z)$ analysis, it was shown that this type of generalization presents a flexible parameter space, which can also solve cosmic tensions, most importantly $H_0$ tension. 
  \item[$\bullet$] When considered together with other evolving dark energy models, SIDE model crosses the phantom divide at a very close redshift value to that of $w_0-w_a$ parametrization. Similarly, both $w(z)$ curves cross $w=0$ axis at similar $z$. Importance of those points; if they are somehow preferred by data for different evolving dark energy models or if those {\it triple} points are coincidental, is to be investigated more in detail in the future.

Strength of a dark energy theory comes from the number of criteria it satisfies. If such theory is a good and consistent fit to every cosmological observation without any tensions, if it has a motivation from particles, fields and/or interactions of those and if it does not lead to any fundamental inconsistencies and/or paradoxes, it should be considered as a valid dark energy candidate. SIDE model seems to check some of those boxes. In the future, we are planning to test its consistency using more diverse observational data and we aim to focus more on the theoretical foundations that are compatible with our phenomenological ansatz. 

\bibliographystyle{unsrt}
\bibliography{sidmbib}

\end{document}